# National Quantum Strategies: A Data-Driven Approach to Understanding the Quantum Ecosystem

Simon Richard Goorney, Emre Aslan, Aleksandrs Baskakovs, Borja Muñoz, Jacob Sherson
Department of Management, Aarhus University, Aarhus Denmark. simon.goorney@mgmt.au.dk, sherson@mgmt.au.dk

*Abstract* - As quantum technologies (QT) move from foundational research toward industrial and societal deployment, national strategies have become critical instruments for shaping the future of this emerging field. In this study, we conduct the first large-scale, data-driven analysis of 62 national quantum strategic documents (QSDs) from 20 countries. Using AI-based natural language processing (topic modeling), we identify 12 topics present in the text, ranging from technical development areas to transversal aspects such as workforce development and governance. Temporal analysis reveals a distinct shift in policy discourse toward applications of QT and commercialisation, and relatively away from basic science. Our findings highlight the increasing diversification of the QT field, and contribute to the growing area of quantum policy studies. We advocate for more AI and data-driven analyses of the quantum ecosystem, to work toward a scalable framework for understanding the technological and societal challenges of the second quantum revolution.

*Index Terms* - Natural Language Processing, Quantum Technology, National Strategies, AI, Governance

## Introduction

Nations around the world are increasingly exploring the societal and economic potential of quantum technologies (QT) across various sectors [1,2]. As the global QT market is projected to reach $106 billion by 2040 [3], interest has been continuously growing in understanding not only the technological development but also its broader societal implications [4-6].

However, as global investments in quantum technologies increase, so too do concerns about emerging "quantum divides"—systemic disparities in access to quantum infrastructure, talent, research capacity, and industrial participation [7]. These divides are not only geographical, separating countries with established quantum research hubs from those with limited technological capacity, but also socio-economic and institutional [7]. Left unaddressed, they risk entrenching inequalities between nations, where leading economies dominate quantum innovation while others rely on external access. At the institutional level, elite universities and research centers benefit from partnerships and funding, while smaller or less-connected institutions can be excluded. In addition, some social groups may benefit disproportionately from access to quantum education, training, and employment, with already privileged populations being favoured. It is therefore essential to understand how different nations worldwide address the development of QT, how these narratives are changing over time, and what the implications are for the development of the quantum industry.

Recent research has investigated these narratives in the context of research grants. For instance, Robertson analysed how the emergence of QT was shaping expectations on Australia by analysing research grants between 2002 and 2020, addressing the technoscientific promises of QT used in these proposals, showcasing how often they framed QT as expected to create economic benefit across different sectors, or as solutions for various societal issues [8]. Recently, one study has looked into how news influences public perception and adaptation of QT [9], and a follow-up investigated the main topics addressed by government policy documents, media articles and business documents [10]. They found that business documents focus more on the technical aspects & application, covering areas of interest related to commercialization potential, while media articles present a more balanced distribution of themes. The authors did note the absence of topics related to political and social implications in business and policy discourses, which can lead to regulatory risks or public mistrust, and they recommended fostering a more transparent engagement when communicating about QT to the public [10].

In this paper, we extend this work in the direction of national strategic documents, leveraging the analytical power of AI (natural language processing) to investigate the overarching themes present in QSDs worldwide. In particular, we may hypothesise that these themes, represented by topics in our analysis, would have changed over time, as QT has advanced in its development and headed toward commercialisation. In order to quantify this change, we will consider the following research questions:

RQ1: What are the dominant themes in quantum strategic documents worldwide?
RQ2: How have these thematic emphases shifted over time?

In addition, a core aim of this work is to understand how AI and data scientific methods can be used to shed light on the quantum ecosystem.

## METHODOLOGY

We investigate RQ1 and RQ2 by assembling a dataset of 62 QSDs, and cleaning and filtering the corpus of text to a total of 12,786 paragraphs. These paragraphs are then classified and clustered into topics. The following section describes the process in detail.

### Data acquisition

The documents were acquired during the period 03/2023-04/2024. Several rounds of interactions with experts were made to form the criteria (see below) for including Quantum Strategic Documents (QSD) in our dataset. The QSDs were collected by conducting an exhaustive Google search. There are 4 formats of documents included in our dataset:

1. Quantum national strategies (16): Documents written by a government or a political union (e.g., EU Commission) that outline future plans for quantum technologies in a given country or region.
2. Government quantum reports (22): Documents written by a government, government agency, or political union (e.g., EU Commission) presenting research findings, workshop results, or policy recommendations
3. Third party quantum reports (23): Documents written by a third party, commissioned by the government to provide insights. These might reflect the outcomes of research, workshops or other forms of consultation.
4. Legal frameworks for QT (1): Legal documents entering a quantum strategy into the law of a given country, signed and enforced by the government.

The resulting dataset contains 62 QSDs from 1999 to 2024 (see Fig 1 and Fig 2.)

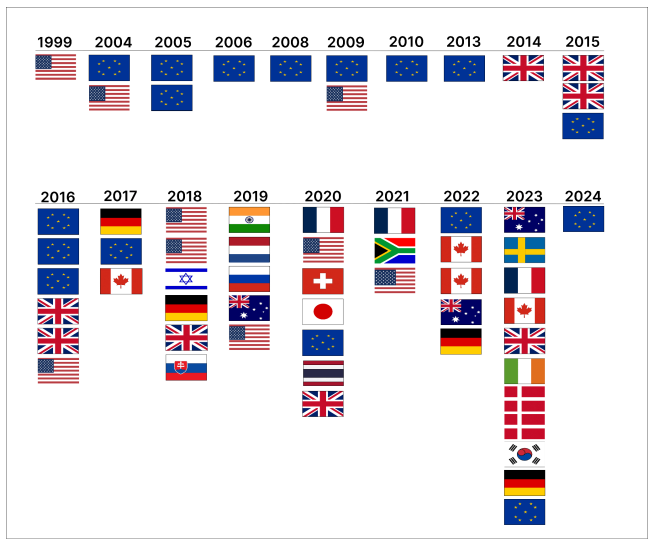

Figure 1. Timeline of 62 documents and their geographical distribution. Includes 16 quantum national strategies, 22 government quantum reports, 23 third party quantum reports and 1 legal framework. QSDs in the dataset date from 1999 (USA) to 2024 (EU)

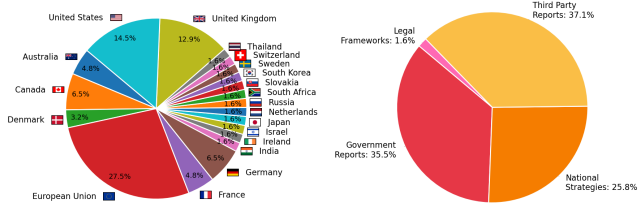

Figure 2. a) Distribution of QSD by country. b) Proportional breakdown of QSD into 4 document formats.

### Data processing

The singular unit of analysis in this study is the individual paragraph within each QSD. These paragraphs vary in content and length, and each paragraph is considered as an independent data point. The rationale for selecting the paragraph as the unit of analysis is twofold. Firstly, a paragraph represents an independent unit of thought as conceived by the authors of QSDs, typically encapsulating a coherent segment of information. Secondly, topic modeling algorithms perform optimally with text of medium length [11]. Paragraphs strike a balance by providing enough context to be meaningful while avoiding the extremes of too little text (a single sentence) or too much text (an entire page). Therefore, using paragraphs as our unit of data ensures both the integrity of the information and the effectiveness of our analytical methods. Paragraphs are extracted using the PyPDF2 library [12], and cleaned using GPT-4 API [13] to remove formatting irregularities and extraneous characters. To verify the quality of the extracted text, a cosine similarity comparison between the raw input text and the processed output was computed. Instances where the cosine similarity score fell below 0.80 were discarded from the dataset. On average, 19% (SD 10%) of pages from the documents were discarded due to low cosine similarity scores. These, however, were mostly text free pages.

The text data was preprocessed through several steps. First, custom stopwords, including country names and random noise words identified during exploratory topic modeling, were removed. All text was then converted to lowercase, and punctuation marks and special characters were eliminated to focus solely on the words themselves. Common stopwords such as "and," "the," and "is" were also removed to reduce noise in the data. The remaining words were lemmatized using the WordNet lemmatizer from NLTK [14], reducing them to their base or dictionary forms for more consistent analysis.

Further cleaning steps were taken to improve the quality and consistency of the dataset. First, paragraphs identified as referring to past events and text concerning initiatives outside of the country of origin were manually flagged and removed to ensure that the analysis focused on the strategic content of the documents rather than retrospective or speculative statements. Second, paragraphs that referred to initiatives being conducted in other countries were also removed to ensure paragraphs were only reflecting the actions of the publishing nation. Rows containing missing text entries were also dropped, and an additional filter was applied to retain only substantial paragraphs—defined as those exceeding 150 characters (50% quantile of text length) or beginning with a bullet point. To remove irrelevant or potentially misleading content, all paragraphs containing URLs were excluded. Following all preprocessing steps, including filtering, lemmatization, custom keyword removal, and categorical refinement, the final dataset consisted of 12,786 paragraphs drawn from 62 national quantum strategy documents. This dataset served as the basis for all subsequent topic modeling and time-series analyses (see Fig 3.)

### Topic modelling

We used the BERTopic model to uncover and characterize the latent themes within QSDs. BERTopic is an innovative topic modeling tool that employs BERT (Bidirectional Encoder Representations from Transformers) embeddings to identify and label topics within a text corpus [14]. BERT, a transformer-based language model, has demonstrated exceptional capabilities in understanding and generating natural language text [11].

### Time-series analysis

To investigate how the thematic composition of QSDs has changed over time, we fitted a series of OLS regression models, one for each of the 12 identified topics. In each model, the dependent variable was the proportion of a given document dedicated to a specific topic (expressed as a percentage), and the independent variable was the release date, scaled in years since the first available document. This allowed us to quantify whether individual topics have become more or less prominent over time across national strategies.

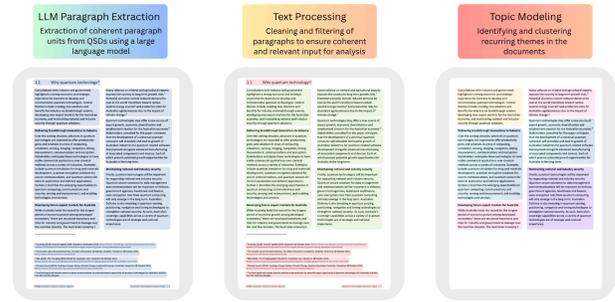

Figure 3: Visualisation of the data processing pipeline. Raw extracted text is separated into paragraphs with the use of an LLM, any text that is not a coherent paragraph is removed, resulting paragraphs are cleaned and classified with the use of a topic model.

### RESULTS

To investigate how the thematic composition of QSDs has changed over time, we analyzed 12 topics extracted across documents released between 1999 and 2024. Each topic's prevalence was quantified as the proportion of a document allocated to it, expressed as a percentage. These proportions were tracked over time using the release dates of the documents, scaled in years since the earliest available publication. Fitting regression models to each topic using release year as the predictor and topic proportion as the outcome, results in a trend showing whether a topic has become more or less prominent over time. Positive slopes suggest increasing interest, while negative slopes indicate diminishing focus. A summary of the regression results for each topic is provided below (Table 1), representing the dominant themes in the 62 QSDs (RQ1).

| Topic | Slope | Intercept | Significance |
|---|---|---|---|
| Atomic and Optical Quantum Platforms | -0.488 | 12.581 | *** |
| Government and National Quantum Programs | 0.599 | -1.32 | * |
| Innovation, Industry, and Investment | 1.448 | -4.612 | *** |
| National and International Research Collaboration | 0.346 | 3.66 | |
| Photonics and Optical Systems | -0.432 | 14.267 | ** |
| Quantum Algorithms and Simulation | -0.212 | 9.81 | |

| Topic | Slope | Intercept | Sig. |
|---|---|---|---|
| Quantum Communication, Networks, and Cryptography | 0.112 | 4.772 | |
| Quantum Computing Hardware, Software, and Applications | -0.349 | 19.634 | |
| Quantum Education and Workforce Development | 0.434 | -1.985 | *** |
| Quantum Physics and Information Theory | -1.346 | 31.861 | *** |
| Quantum Sensing and Applications | 0.184 | 2.906 | |
| Spin and State Control | -0.295 | 8.425 | ** |

Table 1. Summary of OLS regression results for each topic. Reported values include the estimated slope (rate of change in topic prevalence per year), intercept (estimated starting prevalence), and p-value indicating statistical significance. Asterisks denote levels of significance (* $p < 0.05$, ** $p < 0.01$, *** $p < 0.001$).

Each row in Table 1 corresponds to one of the 12 topics and summarizes the estimated linear trend in its prevalence over time. The slope represents the average annual change in topic coverage, measured in percentage points. A positive slope indicates that the topic has, on average, become more prominent in QSDs over time, while a negative slope suggests a decline in attention. The intercept reflects the estimated proportion of content devoted to the topic at the starting point of the timeline (i.e., the earliest document in the dataset).

For example, the topic Atomic and Optical Quantum Platforms has a slope of -0.436 and an intercept of 11.805, indicating that documents at the start of the timeline dedicated approximately 11.8% of their content to this topic, and that this proportion has decreased by about 0.44 percentage points per year over time. Other topics have increased in prevalence over time, and in this section we will address these changes in distribution.

### Temporal evolution of topics

With respect to RQ2, the most significant changes over time are in the topics Atomic and Optical Quantum Platforms (decreasing), Innovation, Industry, and Investment (increasing), Quantum Education and Workforce Development (increasing), and Quantum Physics and Information Theory (decreasing).

### DISCUSSION

The clearest trend we observe appears to be in an overall net shift from the foundational topics to the applied and transversal ones. Quantum Physics and Information Theory and Atomic and Optical Quantum Platforms both have high intercept values representing a significant fraction of the documents in the early years of QIST, however there is now a greater diversity of topics and a movement away from emphasis on basic science in the QSDs. For this reason these topics have decreased in prevalence, although they are of course still essential to the field, they are no longer the core of national strategies.

On the other hand, topics such as Innovation, Industry, and Investment and Quantum Education and Workforce Development have shown an increase. This reflects their essential contribution to industry development and economic success from QT, the primary goal of the QSDs in the present day. The trend becomes clearer when we group the topics into Transversal Topics, Quantum Pillars and Applications, and Basic Science. Transversal topics show a strong increase in prevalence in the documents in comparison to the research and development and technical topics while Quantum Pillars also show a strong increase compared to basic science (Fig 4, Table 2, Table 3).

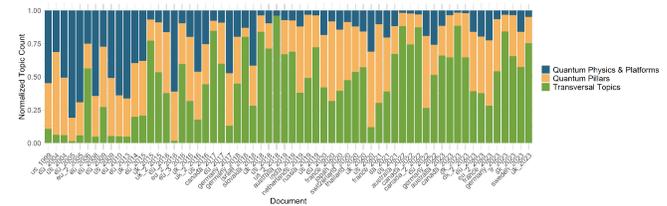

Figure 4: Transversal topics (green) show a clear increase in prevalence over time compared to R&D topics, composed of Quantum Pillars (yellow) and Quantum Physics and Platforms (blue).

| Topic | Slope | Intercept | Significance |
|---|---|---|---|
| Transversal Topics | 2.625 | -1.546 | *** |
| R&D Topics | -2.625 | 101.546 | *** |

Table 2: Comparison between transversal topics (Government and National Quantum Programs, National and International Research Collaboration, Quantum Education and Workforce Development, Quantum Technology: Innovation, Industry, and Investment) and R&D topics (all other topics). The transversal topics show a significant increase in prevalence as the R&D topics show a comparative decrease.

| Topic | Slope | Intercept | Significance |
|---|---|---|---|
| Quantum Physics & | -2.102 | 73.627 | *** |
| Platforms | | | |
| Quantum Pillars | 2.102 | 26.373 | *** |

Table 3: Comparison between Quantum Physics & Platforms (Atomic and Optical Quantum Systems, Spin and State Control, Photonics and Optical Systems, Quantum Physics and Information Theory) and Quantum Pillars (Quantum Computing Hardware, Software, and Applications, Quantum Sensing and Applications, Quantum Communication, Networks, and Cryptography, Qubits and Error Correction in Quantum Computing, Quantum Algorithms and Simulation) The quantum pillars and applications are increasing in prevalence compared to basic science, which show a comparative decrease.

These results tell us something about the state of the quantum industry. The clearest finding is that QT is moving from an era of fundamental research and scientific progress to applications, commercialisation, and industry development. For nations worldwide, this requires not only scientific sovereignty, but also education and workforce development, to ensure that talent is available to take up roles in the developing quantum industry and that skills are not a bottleneck in its development. This is demonstrated by the significant increase in prevalence of this topic over time. Industry development is also being prioritised in the QSDs, with specific actions such as accelerator programs, seed funding, and support for startups and SMEs a key aspect of the documents in recent years.

Looking back to the first document from 1999, which was the NSF report on QIST (quantum information science and technology), there it was described as "The birth of a new science". In the present day, we are firmly in the centre of this new science, and we are as much in engineering and commercialisation as we are in scientific progress. This is reflected in the topical analysis, but it is also clear in conferences, journals, and scientific discussions worldwide, that QIST is making its way toward widespread applications. Large national and international infrastructure, such as the EU's EuroQCI, are well under way. This research demonstrates, in a data scientific manner, that we are indeed heading toward this future.

## Limitations

Whilst this work draws on a large corpus of data, here are some limitations to bear in mind. With regard to the analysis, the simple linear regression is appropriate given the data constraints, as the dataset consists of 62 documents distributed unevenly over a 25-year period, with each document treated as a unique time point. While more flexible models were considered - such as polynomial regression to capture potential non-linearities (e.g., U-shaped trends) - these produced highly uncertain estimates due to the small sample size. As a result, linear regression was selected as a more stable and interpretable approximation, even though it may fail to capture more complex temporal dynamics, such as specific new breakthroughs or topical foci in particular time periods. A further limitation concerns the lack of modeling for country-specific effects. Ideally, one would account for national differences by including the country as a random effect in a multilevel model, to reflect how different technological capacities, research infrastructures, or policy environments shape topic emphasis. However, the dataset contains too few repeated observations per country to support such a structure. Such an analysis may be possible in another decade, when more countries have multiple generations of QSD to be included. Finally, as Google translate was used for translating some of the documents, the accuracy of information in these documents may have reduced. However, one should bear in mind that topic modelling clusters keywords, and therefore translations may still contain the same lexical content even if they do not read perfectly to a non-native speaker, and this has been demonstrated in research [15]. Therefore, this is likely to be a rather small effect with minimal loss of accuracy.

## Conclusion

This study provides the first large-scale, data-driven analysis of national quantum strategies (QSDs) using natural language processing, offering a systematic view of how the discourse around quantum technologies has evolved globally over time. By applying topic modeling to over 12,000 paragraphs from 62 QSDs, we identified key thematic trends and shifts in national priorities. Our findings reveal a notable transition from foundational topics—such as quantum physics and photonics—toward applied, system-level themes including quantum computing hardware, industry growth, and workforce development. This transition underscores a broader maturation of the quantum technology ecosystem, where policy documents increasingly emphasize commercialization, infrastructure-building, and societal impact. The rise in prominence of transversal topics, such as education and national program coordination, further suggests that governments are not only investing in technical capabilities but also in the enabling conditions necessary for sustained growth and competitiveness.

This work contributes to both the empirical understanding of quantum policy discourse and the methodological toolkit available to science and technology studies. Our use of paragraph-level BERTopic modeling combined with time-series regression analysis allows for a fine-grained view of strategic shifts, complementing more qualitative interpretations of national agendas. These insights are particularly relevant as countries position themselves in the global quantum landscape, and their choice of topics gives insight into their strategies. In the future, we encourage new research to integrate our approach with sentiment and narrative analysis to better understand how ambition, risk, and national identity are framed in these strategies. An immediate follow up we are working on is comparing between nations in which areas of QT they are working toward supremacy, and in which areas they are choosing to work together. The QSDs are a wealth of information which may help to further understand how nations worldwide are working on

An additional future consideration that we wish to highlight here, is that we must now understand how these top-down changes in emphasis of national governments worldwide are filtering into the industry. How many jobs are being created by these national ambitions for commercialisation? And in what fields are they, what skills do new graduates need? Initial research in this direction has been undertaken by several authors [16-18], with a number of articles finding that the industry remains relatively nascent, with few non technical jobs available. However, graduates from a wide variety of backgrounds, including those with PhDs but also Bachelor's and Master's, are finding positions - which is a positive sign for increasing diversification and commercialisation. As quantum technologies approach broader deployment, ensuring that national strategies are anticipatory, and societally informed will be essential to shaping a responsible quantum future and successful industry worldwide.

## Funding Acknowledgement

This work was supported by the Horizon 2020 project QUCATS under Grant Agreement ID 101070193 and Digital Europe project QTIndu under Grant Agreement ID 101100757.